\documentstyle[amssymb,preprint,aps]{revtex}

\begin{document}
\author{M. Jaime$^a$, H. T. Hardner$^a$, M.B. Salamon$^{a,b}$, M. Rubinstein$^c,$ P.
Dorsey$^c,$ and D. Emin$^d$}
\address{$^a$Department of Physics and Materials Research Laboratory\\
University of Illinois at Urbana-Champaign. 1110 W. Green Street, \\
Urbana IL, 61801-3080\\
$^b$Center for Material Science, Los Alamos National Laboratory, \\
Los Alamos, NM 87545\\
$^c$U.S. Naval Research Laboratory, Washington, D.C. 20375-5000\\
$^d$Sandia National Laboratories,Albuquerque, NM 87185}
\title{Hall-Effect Sign Anomaly and Small-Polaronic Conduction in (La$_{1-x}$Gd$_x$)%
$_{0.67}$Ca$_{0.33}$MnO$_3$}
\date{9/12/96}
\maketitle

\begin{abstract}
The Hall coefficient of Gd-doped La$_{2/3}$Ca$_{1/3}$MnO$_3$ exhibits
Arrhenius behavior over a temperature range from $2$ $T_c$ to $4$ $T_c$,
with an activation energy very close to $2/3$ that of the electrical
conductivity. Although both the doping level and thermoelectric coefficient
indicate hole-like conduction, the Hall coefficient is electron-like. This
unusual result provides strong evidence in favor of small-polaronic
conduction in the paramagnetic regime of the manganites.

\end{abstract}


A recent resurgence of interest in the transport properties of doped
lanthanum manganites has resulted in the realization that electron-lattice
interactions play a significant role. Studies of the archetypal system, (La$%
_{1-x}$R$_x$)$_{1-c}$Ca$_c$MnO$_3$, have demonstrated that the ferromagnetic
transition temperature $T_c$ (and with it, the metal-semiconductor
transition) are suppressed by the addition of rare-earth ions R whose
smaller size further closes the Mn-O-Mn bond angle.\cite{font,hwang} The
temperature dependence of the resistivity above $T_c$ is remarkably
unaffected by rare-earth substitutions, following a universal
semiconductor-like curve. This supports the view that replacement of a
trivalent rare-earth ion by Ca introduces a hole that is presumably
associated with the Mn{\em \ }$e_g$ state. Theoretical attempts \cite
{millis,roder} to describe the large changes in resistivity and their
sensitivity to magnetic fields in the vicinity of $T_c$ in the context of
the double exchange model have led to the conclusion that strong
electron-lattice effects are essential, and that the transition shares
aspects of ``polaron collapse'' such as occurs in EuO. \cite{emin} \ The
conductivity in the high temperature regime should be dominated by the
hopping motion of self-trapped, small polarons. Indeed, quite recent
experiments have shown the importance of electron-phonon interactions in CMR
via the oxygen isotope effect. \cite{zhao}

A stable polaron in an ionic solid may be either a large (multisite) polaron
that moves itinerantly, or a small (single-site) polaron that moves with a
low ($<<$ $1$ cm$^2/$Vs) thermally assisted mobility. \cite{e&h} In the
single-site limit, the self-trapped carrier's energy is taken to depend only
linearly on the displacement of atoms from their carrier-free positions.
Consequently, the characteristic energy of the Seebeck coefficient E$_s$ is
significantly smaller than the activation energy of the electrical
conductivity E$_\sigma $; that E$_s<<$ E$_\sigma $ was demonstrated in
earlier work, \cite{jaime} and taken as evidence of small-polaronic motion. E%
$_s$ measures the chemical potential of the self-trapped polaron.

Perhaps the most distinctive property of steady-state small-polaronic
transport is its Hall mobility $\mu _H$. The activation energy of the Hall
mobility is calculated to be always less than that for drift mobility E$_d.$%
\ The simplest model predicts $\approx $E$_d/3,$ and this has been observed
in, for example, oxygen-deficient LiNbO$_3.$ \cite{nagels} The sign of the
Hall effect for small polaron hopping can be ``anomalous.'' A small polaron
based on an electron can be deflected in a magnetic field as if it were
positively charged and, conversely, a hole-based polaron can be deflected in
the sense of a free electron. As first pointed out by Friedman and Holstein 
\cite{fh}, the Hall effect in hopping conduction arises from interference
effects of nearest neighbor hops along paths that define an Aharonov-Bohm
loop. Sign anomalies arise when the loops involve an odd number of sites.%
\cite{emin1} In this Letter, we report the first measurement of the
high-temperature Hall coefficient of manganite samples, finding that it
exhibits Arrhenius behavior and a sign anomaly relative to both the nominal
doping and the thermoelectric power. The results are discussed in terms of
an extension of the Emin-Holstein (EH) theory \cite{e&h} of the Hall
mobility in the adiabatic limit.

We exploit the sensitivity of these materials to rare-earth substitutions to
lower the transition temperature from $\sim 260\;$K at $x=0$ to $\sim 130$
K, thereby extending the accessible temperature range to $\simeq 4$ $T_c.$
The samples used in this study were laser ablated from ceramic targets and
deposited on LaAlO$_3$ substrates as described previously. \cite{jaime,white}
The ceramic target material (R = Gd and $x=0.25)$ has a resistivity maximum
near $100$ K while the thermoelectric power drops abruptly to metallic
values below $90K$. \cite{white} The effective rare-earth site radius is $%
\left\langle r_A\right\rangle =$ $0.113$ nm and, indeed, the properties of
this sample are very similar to Y-substitute samples with the same $%
\left\langle r_A\right\rangle $ value. \cite{font} The laser deposited films
show somewhat higher transition temperatures: the resistivity maximum is at $%
140$ K and the thermoelectric power becomes metal-like below $130$ K. X-ray
data indicate that the in-plane lattice parameter of the laser ablated film
is larger than that of the ceramic, presumably due to stress induced by the
substrate. Fig. 1 shows resistivity and thermopower data for the $130$ K
sample along with resistivity data for similar samples with $x=0$ and $%
x=0.5. $ We focus here on the $130$ K sample, for which $\rho $ drops by a
factor $160$ between $0$ and $8$ Tesla at $142$ K at which field the peak in
the thermopower is almost completely suppressed.

The most rapid motion of a small polaron occurs when the carrier hops each
time the configuration of vibrating atoms in an adjacent site coincides with
that in the occupied site. This regime is termed adiabatic with a
conductivity given by $\sigma =\sigma _o\exp (-E_\sigma /$k$_B$T$),$ where 
\begin{equation}
\sigma _o=g_de^2\nu _0/ak_BT.  \label{eq1}
\end{equation}
Here, $\nu _0$ is a characteristic frequency and $a$ is the jump distance,
which we take to be the Mn-Mn spacing, $0.39\;$nm. The factor $g_d$ depends
on hopping geometry and has the value $g_d=3/2$ for the triangular lattice
treated by EH and $g_d=1$ for nearest-neighbor hopping on a square planar
lattice. A signature of the adiabatic limit is that the prefactor $\sigma _o$
approaches$\;e^2/\hbar a\simeq 7000\;$($\Omega $ cm)$^{-1}$ when $h\nu
_0\simeq k_BT$. If it is much smaller, the motion is termed nonadiabatic,
and the prefactor contains an additional factor of T$^{-1/2}.$ In Fig. 2 we
plot -$\log (\sigma $T$)$ and -$\log (\sigma $T$^{3/2})$, the adiabatic and
nonadiabatic limits respectively, vs 1/T$.$ Close to $T_c$ there are
significant deviations from Arrhenius behavior, as has been noted
previously. \cite{white,emin2} At higher temperatures, there is no
significant difference between adiabatic and nonadiabatic fits; the fitting
parameters are given in Table 1 along with those for Gd concentration $x=0$
and $x=0.5$. The adiabatic prefactor for the $x=0.25$ sample is $\approx
6000\;\Omega ^{-1}$cm$^{-1}$ at $300$ K, confirming that the hopping
processes are adiabatic. The characteristic frequency $\nu _0$ in Eq.\ (1)
is of order $10^{14}/g_d$ Hz. Evidence, discussed below, implies that $g_d$
is significantly larger than the value for pure nearest-neighbor hopping ($%
g_d=1$). This reduces $\nu _0$ to a value characteristic of optical phonons
in transition-metal oxides.

Sections of these specimens were patterned by conventional lithographic
methods into a five-terminal Hall geometry. Hall experiments were carried
out in a high-temperature insert constructed for use at the $20$ Tesla
superconducting magnet at the National High Magnetic Field Laboratory (Los
Alamos, NM). Figure 3 shows the raw data (transverse and longitudinal
voltages) obtained in for $x=0.25$ at $462$ K. Although the sample
lithography resulted in negligible zero field transverse voltage in the
metallic state, a transverse magnetoresistive signal is apparent above $T_c;$
see Fig. 3. This signal peaks around $T_c$ and then decreases with
increasing temperature. Clearly, this indicates inhomogeneous current paths
in the proximity of the metal-insulator transition, possibly due to local
relaxation of epitaxial strain and resulting local variations in $T_c$.
Consequently, we did not attempt to follow the behavior of the Hall constant
below $\sim 2T_c.$ The transverse voltage data taken while sweeping the
field from $-16$ Tesla to $+16$ Tesla, and that taken while sweeping back to 
$-16$ Tesla, were each fit to a second-order polynomial with the term linear
in field attributed to the Hall effect. We verified in each case that the
longitudinal magnetoresistance is completely symmetric in field. Fig. 4
shows the Hall coefficient derived from the linear term. Several points for
the $x=0$ film are included. Due to the much higher $T_c$ of that sample,
extraction of the Hall contribution leads to greater uncertainty. The data
are, however, consistent with the Gd-substituted film. The line through the
data points is an Arrhenius fit, giving the expression R$_H=-(3.8\times
10^{-11}\;$m$^3/$C$)\exp (91\;$meV$/$k$_B$T$).$ Note that the sign is
negative, even though divalent dopants should introduce holes. That the
Seebeck coefficient approaches a negative value at high temperatures has
been attributed in part to the reduction in spin entropy produced when a
hole converts a Mn$^{3+}$ ion to Mn$^{4+}.$ \cite{jaime}

Detailed expressions for the Hall effect in the adiabatic limit have been
calculated by EH \cite{e&h} for the hopping of electrons with positive
transfer integral $J$ on a triangular lattice, and results in a normal
(electron-like)\ Hall coefficient. However, the sign of both the carrier and
the transfer integral changes for hole conduction, leaving the sign of the
Hall coefficient electron-like, and therefore anomalous. \cite{fh,emin1}
However, no anomaly would arise if the hopping involves 4-sided loops with
vertices on nearest-neighbor Mn atoms. A sign anomaly, then, implies that
hopping involves odd-membered Aharonov-Bohm loops. Such processes arise when
next-nearest neighbor (nnn) transfer processes across cell face diagonals
are permitted. If the Mn-O-Mn bonds were strictly colinear, the former
processes would be disallowed by symmetry. However, the bond angles are
substantially less than $180^{\circ }$, implying the presence of $\pi $-bond
admixture, and opening a channel for diagonal hops. We have extended the
triangular-lattice calculation of EH to the situation in which a hole on a
Mn ion can hop to any of its four nearest neighbors in the plane normal to
the applied field with transfer matrix element J $<0$ {\it and }to its four
next-nearest-neighbors (nnn) with a reduced transfer energy $\gamma $J$.$ We
must also consider the effect of these diagonal hops (plus those in the
plane containing both electric and magnetic fields) on the conductivity
prefactor, Eq. (1). The Hall coefficient can be written as R$_H=$R$_H^o($T$%
)\exp (2E_\sigma /3k_BT),$ with 
\begin{equation}
R_H^o=-\frac{g_H}{g_d}\frac{F(|J|/k_BT)}{ne}\exp \left[ -[\epsilon
_0+(4|J|-E_s)/3]/k_BT\right] ;  \label{eq2}
\end{equation}
EH\ found that the factor $g_H=1/2$ for three-site hopping on a triangular
lattice. In Eq.(2) we have expressed the carrier-density as $n\exp
(-E_s/k_BT),$ where $E_s$ is estimated to be 8 meV from the thermopower
data. The quantity $\epsilon _0$ is the $J-$dependent portion of a carrier's
energy achieved when the local electronic energies of the three sites
involved in an Aharonv-Bohm loop are equal. For the problem considered by
EH, an electron hopping within a lattice composed of equilateral triangles, $%
\epsilon _0=-2|J|,$ and $g_H/g_d=1/3.$ Within the domain of validity of EH,
the temperature dependence of $R_H$ arises primarily from the factor $\exp
(2E_\sigma /3k_BT)$ when $E_\sigma >>E_s.$ For holes hopping within a cubic
lattice in which three-legged Aharonov-Bohm loops include face-diagonal
transfer, we find $\epsilon _o=-|J|(\sqrt{8+\gamma ^2}-\gamma )/2$. In
particular, $\epsilon _o$ varies from -$\sqrt{2}|J|$ to $-|J|$ as $\gamma $
increases from zero to unity \cite{woods}, and the temperature dependence of 
$R_H$ remains dominated by the factor $\exp (2E_\sigma /3k_BT).$ Indeed, the
energy characterizing the exponential rise of the Hall coefficient that we
observe, E$_{Hall}=91\pm 5\;$meV, is about 2/3 the measured conductivity
activation energy, E$_{Hall}/$E$_\sigma =0.64\pm 0.03,$ in excellent
agreement with theory.

The geometrical factor $g_d$ depends on the ratio of the probability $%
P_{nnn} $ of nnn hops to $P_{nn},$ that of nn hops, through $%
g_d=(1+4P_{nnn}/P_{nn}).$ If these probabilities are comparable ($\gamma
\sim 1)$ $g_d=5,$ $g_H=2/5$ and the exponential factor in Eq.\ (2) becomes $%
\exp [(E_s-|J|)/3k_BT]\simeq 1.$ In the regime $|J|\gtrsim k_BT,$ the
function $F(|J|/k_BT)$ is relatively constant with a value $\approx 0.2,$
and we find $R_H^o\simeq -0.02/ne=-3.8\times 10^{-11}$m$^3$/C. This yields
an estimated carrier density $n=3.3\times 10^{27}$m$^{-3},$ quite close to
the nominal level of $5.6\times 10^{27}$ m$^{-3}.$ Diagonal hopping also
reduces the value of the attempt frequency $\nu _0$ required to fit the
conductivity prefactor to $\approx 2\times 10^{13}$ Hz, as noted above.

In conclusion, we have measured the high-temperature Hall coefficient in
manganite films and found that its temperature dependence is consistent with
small-polaron charge carriers that move by hopping. Further, the magnitude
of the conductivity prefactor indicates that the carrier motion is
adiabatic. Finally, the sign anomaly in the Hall effect implies that small
polarons hop not only among near-neighbor sites (making Aharonov-Bohm loops
with an even number of legs) but must have a significant probability of
traversing Hall-effect loops with odd numbers of legs. As such, the results
indicate the occurrence of significant nnn transfer across face diagonals,
and therefore a crucial role for deviations of the Mn-O-Mn bond angle from $%
180^{\circ }$. An interesting possibility, that may also relate to unusual
high-temperature values observed for the Seebeck coefficient \cite{jaime},
is that transport is a type of impurity conduction in which carriers remain
adjacent to divalent cation dopants (i.e. Ca ions). The local distortions
associated with the presence of the impurity may also increase the admixture
of $\pi $-bonds, and enhance diagonal hopping.

We are pleased to acknowledge useful discussions with J.B. Goodenough and
C.P. Flynn, and the assistance of A. Lacerda with high fields experiments.
This work was supported by the Department of Energy, office of Basic Energy
Sciences through Grant No. DEFG02-91ER45439 at the University of Illinois
and under contract DE-AC04-94AL85000 at Sandia National Laboratory. MBS\ was
supported, in part, as a Matthias Fellow at Los Alamos National Laboratory.

\begin{table}[tbp] \centering%
\begin{tabular}{|l|l|l|l|l|}
\hline
(La$_{1-x}$Gd$_x$)$_{2/3}$Ca$_{1/3}$MnO$_3$ & E$_\sigma ($m$e$V$)$ & $\sigma
_o(\Omega ^{-1}$cm$^{-1})$ & E$_H($m$e$V$)$ & R$_H^o(10^{-10}$ m$^3/$C$)$ \\ 
\hline\hline
sample A ($x=0$) &  &  &  &  \\ \hline
Nonadiabatic limit & $102$ & $2.5\times 10^7$ K$^{3/2}/$T$^{3/2}$ & $\sim 69$
&  \\ 
Adiabatic limit & $85$ & $7.7\times 10^5$ K$/$T & $\sim 59$ &  \\ 
\hline\hline
sample B ($x=0.25$) &  &  &  &  \\ \hline
Nonadiabatic limit & $158$ & $5.1\times 10^7$ K$^{3/2}/$T$^{3/2}$ & $112$ & (%
$4.8\times 10^{-4}\;$K$^{-1})$T \\ 
Adiabatic limit & $145$ & $1.8\times 10^6$ K$/$T & $91$ & $0.38$ \\ 
\hline\hline
sample C ($x=0.5$) &  &  &  &  \\ \hline
Nonadiabatic limit & $157$ & $2.3\times 10^7$ K$^{3/2}/$T$^{3/2}$ &  &  \\ 
Adiabatic limit & $146$ & $8.9\times 10^5$ K$/$T &  &  \\ \hline
\end{tabular}
\caption{Parameters from adiabatic and non-adiabatic fits to the resistivity and Hall data
\label{t1}}%
\end{table}%

\begin{figure}[tbp]
\caption{ a) The resistivity vs temperature for the $x = 0.25$ sample, for
magnetic fields ($\blacksquare $) H = 0 and ($\square $) H = 8 Tesla. 
Data for $x= 0$ and $x=0.5$ are shown for comparison
b) The thermoelectric power for $x = 0.25$ in the same range of temperatures 
and fields showing a saturated $S$(H) at temperatures T $<T_c$.}
\label{fig.1}
\end{figure}
%

\begin{figure}[tbp]
\caption{The resistivity of the $x=0.25$ sample plotted in the adiabatic
(squares) and non-adiabatic (thick line) limits. The fine line is a linear
fit indicating an activation energy of 145 m$e$V in the high temperature
adiabatic regime. }
\label{fig.2}
\end{figure}
%

\begin{figure}[tbp]
\caption {Raw Hall data in $x=0.25$ sample at T =462 K, showing an odd
contribution in the transverse voltage V$_{xy}$ not present in
the longitudinal voltage V$_{xx}$.}
\label{fig.3}
\end{figure}
%

%
\begin{figure}[tbp]
\caption{The magnitude of  the Hall coefficient -R$_H$ {\em vs.} temperature showing 
the values obtained ramping the magnetic field up (solid triangles) and down 
(solid nablas) in the $x=0.25$ sample. Open squares  correspond to $x=0$, and
have large error bars. The dashed line is an Arrhenius fit.
Inset: The natural logarithm of the Hall coefficient (average of up and down field values)
 for $x=0.25$ vs 1000/T. The solid line is a linear fit giving 91 m$e$V for the activation
energy.}
\label{fig.4}
\end{figure}

%


\begin{references}
\bibitem{font}  J. Fontcuberta, B. Martinez, A. Seffar, S. Pi\~{n}ol, J.L.
Garc\'{\i }a-Mu\~{n}oz and X. Obradors, Phys. Rev. Lett. {\bf 76}, 1122
(1996); J. Fontcuberta, et. al. Appl. Phys. Lett. {\bf 68}, 2288\ (1995).

\bibitem{hwang}  H.Y. Hwang, S-W. Sheong, P.G. Radaelli, M. Marezio, and B.
Batlogg, Phys. Rev. Lett. {\bf 75}, 914(1995); Note: the authors appear to
have used incorrect coordination numbers in calculating $\left\langle
r_A\right\rangle .$

\bibitem{millis}  A.J. Millis, P.B. Littlewood, and B.I. Shraiman, Phys.
Rev. Lett{\it .} {\bf 74}, 5144\ (1995); A.J. Millis, et al. (preprint,
1996).

\bibitem{roder}  H. R\"{o}der, J. Zang, and A.R. Bishop, Phys. Rev. Lett{\it %
.} {\bf 76}, 1356 (1996).

\bibitem{emin}  D. Emin, M. Hillery, and N-L. H. Liu, Phys. Rev. B. {\bf 33}%
, 2933 (1986).

\bibitem{zhao}  G. Zhao, K. Conder, H. Keller, K. A. M\"{u}ller{\it , }Nature%
{\it \ }{\bf 381}, 676 (1996).

\bibitem{e&h}  D. Emin and T. Holstein, Ann. Phys. (N.Y.) {\bf 53}, 439
(1969).

\bibitem{jaime}  M. Jaime, M.B. Salamon, K. Pettit, M. Rubinstein, R.E.
Treece, J.S. Horowitz, and D.B. Chrisey, Appl. Phys. Lett. {\bf 68,} 1576
(1996); M. Jaime, et al. Phys. Rev. B{\it \ }(in press).

\bibitem{nagels}  P. Nagels in {\it The Hall Effect and its Applications},
Edited by C.L. Chien and C.R. Westgate (Plenum, NY 1980) p. 253

\bibitem{fh}  L. Friedman and T. Holstein, Ann. Phys.\ (N.Y.) {\bf 21}, 494
(1963).

\bibitem{emin1}  D. Emin, Phil. Mag{\it .} {\bf 35}, 1189 (1977).

\bibitem{white}  P. White, M. Jaime, M.B. Salamon and M. Rubinstein, Bull.
Am. Phys. Soc{\it .} {\bf 41}, 116 (1996).

\bibitem{emin2}  D. Emin, Phys. Rev. Lett. {\bf 32}, 303 (1974).

\bibitem{woods}  C. Wood and D. Emin, Phys. Rev. B {\bf 29}, 4582 (1984).
\end{references}
\end{document}